\documentclass[aps,prd,twocolumn,amsmath,amssymb]{revtex4}
\usepackage{natbib}
\usepackage{graphicx}% Include figure files
\usepackage{dcolumn}% Align table columns on decimal point
\usepackage{bm}% bold math

\begin{document} 

\title{Modified Tsallis and Weibull Distributions for multiplicities in $e^{+}e^{-}$ collisions}
\author{S. Sharma}
\author{M. Kaur}
\email{manjit@pu.ac.in} 
\author{S. Thakur}%
\affiliation{Department
of Physics, Panjab University, Chandigarh -160 014, India.}

\date{\today}% It is always \today, today,
             %  but any date may be explicitly specified

\begin{abstract}
Multiplicity distributions of charged particles produced in the $e^{+}e^{-}$ collisions at energies ranging from 14 to 91 GeV are studied using Tsallis $q$-statistics and the recently proposed Weibull distribution functions, in both restricted rapidity windows as well as in full phase space. It is shown that Tsallis $q$-statistics explains the data excellently in all rapidity ranges while the Weibull distribution fails to reproduce the data in full phase space. Modifications to the distributions are proposed to establish manifold improvements in the fitting of the data.
\end{abstract}  

\maketitle

\section{Introduction}

Particle collisions at very high energies produce quark-quark, quark-gluon and gluon-gluon interactions which result in the production of a multitude of elementary particles. Several of these particles being mesons, baryons and leptons. This particle production is described in terms of several theoretical and phenomenological models derived from  quantum chromodynamics. In high energy collisions, particles are visualised to be produced in an intense environment at very high temperature and the formation of quark-gluon plasma in the interactions of quarks and gluons. Thus several models use laws of fluid mechanics, statistical mechanics, thermodynamics, hydrodynamics etc. to describe the particle production. These models have been intriguingly successful. Present day high energy experiments, include several layers of detectors capable of detecting and recording the particles, both neutral and charged, produced in collisions and study the charged particle multipliities, in particular. The distributions are then matched with predictions from various phenomenological models to understand the production mechanism. Concepts from ensemble theory in statistical mechanics have been used to develop models which include statistical fluctuations as an important source of information. Distributions derived from statistics such as Poisson distributions, negative binomial distribution \cite{NB1}, \cite{NB2}, KNO scaling law \cite{KNO} etc. have also played an important role in the understanding of multiplicity distributions, particularly at low energies. The multiplicity, however does not increase linearly with energy and hence at high energies, several new distributions have been proposed. Some of these include  MNBD \cite{MNBD}, KW \cite{KW}, Tsallis \cite{TS1}\cite{TS2}, Gamma \cite{Gam1} \cite{Gam2}, Log-normal \cite{LN}, Weibull \cite{Wei} etc. distributions. In the Tsallis $q$-statistics, standard statistical mechanics becomes non-extensive. The non-extensive property of the entropy is quantified in terms of a parameter $q$ which is shown to be more than unity under this assumption. The non-extensivity property of entropy is used to derive the multiplicity distribution. Weibull distribution is another statistical distribution which has been studied recently \cite{Wei} to describe multiplicity distributions in $e^{+}e^{-}$ by S. Dash et al.   
In the present study, our focus is to investigate the multiplicity distributions, mostly in restricted rapidity windows, at different energies and to study the characteristic properties of charged particle production in $e^{+}e^{-}$ collisions. In one of our earlier papers \cite{SS}, we used Tsallis distributions to fit $e^{+}e^{-}$ data from 34.8 to 206 GeV of energy in the full phase space and modified the Tsalis distribution to obtain the best fits as compared to several other distributions. In this paper we will limit ourselves to comparing the distributions using Tsallis $q$-statistics with the Weibull distribution in both restricted rapidity regions as well as in the full phase space to understand the constraints for the models used. We also propose a modification to improve the comparison between the predicted and the experimental values. Very interesting observations are made.
 
\section{Charged Multiplicity Distributions}
Charged particle multiplicity is defined as the average number of charged particles produced in a collision $<n_{ch} >= \frac{\sum {n}}{N}$, where N is the total number of interactions. Angles at which these particles are produced, are measured in terms of Rapidity defined as $y = -ln\frac{E+P_{L}}{E-P_{L}}$ where E is the particle energy and $P_{L}$ is the longitudinal momentum. We briefly outline the distributions used to study the multiplicity distributions;

\subsection{Tsallis distribution}
Tsallis statistics deals with entropy in the usual Boltzman-Gibbs thermo-statistics modified by introducing 
$q$-parameter and is defined as;

\begin{equation}
S = \frac{1-\sum_{a}P_{a}^q}{q-1}\, \label{one}
\end{equation}
where  $P_a$  is the probability associated with microstate $a$ and sum of the probabilities over all microstates is normalized to one;
   $\sum_{a}P_{a}=1$.

In Tsallis $q$- statistics probability is calculated using the partition function Z. Tsallis entropy is defined as
 \begin{equation}
 S_q(A,B)= S_A + S_B+ (1-q)S_{A}S_{B}
 \end{equation}
where $q$ is entropic index with value, $q>1$ and $1-q$ measures the departure of entropy from its extensive behaviour.

Probability is defined in terms of partition function as 
 
\begin{equation}
P_N = \frac{Z^{N}_q}{Z}
\end{equation}
where Z represents  the total partition function and $Z^{N}_q$ represents partition function at a particular multiplicity.

For N particles, partition function can be written as
 
\begin{equation}
Z(\beta,\mu,V) = \sum(\frac{1}{N!})(nV-nv_{0}N)^{N}
\end{equation}
$n$ represents the gas density, V is the volume of the system and $v_{0}$ is the excluded volume.  
 \begin{equation}
 \bar{N}=Vn[1 + (q-1)\lambda(V n\lambda-1)- 2v_0n], 
 \end{equation} 
 $\bar{N}$ represents the average number of particles. Details of the Tsallis distribution and how to find the probability distribution can be obtained from \cite{TS2}. In one of our earlier papers, we have analysed the $e^{+}e^{-}$ interactions at various energies for full phase space data and described the procedure in detail in reference \cite{SS}.
 
\subsection{Modified Tsallis distribution}
 
In our earlier paper \cite{SS}, we proposed to modify the multiplicity distribution in terms of two components; one due to multiplicity in 2-jet events and another due to multi-jet events. We then calculated the probability function from the weighted superposition of Tsallis distributions of these two components as described below. 

\begin{multline}
P_{N}(\alpha:\bar{n_1},V_1,v_{01},q_1:\bar{n_2},V_2,v_{02},q_2)= \\
\alpha P_{N}(\bar{n_1},V_1,v_{01},q_1) +\\
(1-\alpha)P_{N}(\bar{n_2},V_2,v_{02},q_2)
\end{multline} 

where $\alpha$ is a weight factor which gives 2-jet fraction from the total events and is determined from a jet finding algorithm.
 
\subsection{Weibull distribution} 

 Weibull distribution is a continuous probability distribution which can take many shapes. It can also be fitted to non-symmetrical data. 
 
The probability density function of a Weibull random variable is;
\begin{multline}
 P_N(N,\lambda,k) = \Bigg\lbrace\frac{k}{\lambda} (\frac{N}{\lambda})^{(k-1)} exp^{-(\frac{N}{\lambda})^{ k}} \hspace{0.8cm} N \geq 0 \\                                   
0  \hspace{3.6cm}  N < 0 
\end{multline} 
 
The standard Weibull has characteristic value $\lambda >0$ and shape parameter $k>0$ for its two parameters.The two parameters for the distribution are related to the mean of function, as 
\begin{equation}
\bar{N} = \lambda \Gamma(1+1/k)
\end{equation}

\subsection{Modified Weibull distribution}
Modified Weibull distribution has been obtained by the weighted superposition of two Weibull distributions and used to produce the multiplicity distribution. We convolute the weighted distributions due to 2-jet component and multi-jet components of the events, as below; 

\begin{multline}
P_{N}(\alpha:N_1,\lambda_1,k_1;N_2,\lambda_2,k_2)=\\
\alpha P_{N}(N_1,\lambda_1,k_1) + \\
(1-\alpha)P_{N}(N_2,\lambda_2,k_2) 
\end{multline}

where $\alpha$ is the weight factor for 2-jet fraction from the total events and the remaining $1-\alpha$ is the multi-jet fraction. $\alpha$ is calculated from the DURHAM jet algorithm, as discussed in the next section.

\section{Analysis on Experimental data \& Results}

 Experimental data on $e^{+}e^{-}$ collisions at different collision energies from different experiments are analysed. The data used are from the experiments, TASSO  \cite{TASSO}, ALEPH \cite{ALEPH} and DELPHI \cite{DELPHI} at $\sqrt{s}$ = 14, 22, 34.8, 44, 91 GeV and from the restricted rapidity windows of $|y|< 0.5, 1.0, 1.5, 2.0$. Charged particle multiplicity distributions in terms of probability distributions as $P_n =\frac{\sigma_n}{\sigma_{tot}}$, where $\sigma_n$ is the cross section for multiplicity $n$ and $\sigma_{tot}$ represents the total cross section of the interaction at center of mass energy $\sqrt{s}$. Experimentally this probability can be obtained using number of charged particles produced at specific multiplicity, $n$ and total number of particles, $N_{tot}$ produced in whole process, $P_{n}= \frac{n}{N_{tot}}$. 
The experimental distributions are fitted with the predictions from Tsallis $q$-statistics and the Weibull distribution as described in the following approaches.
 
\subsection{Approach I}
The probability distributions using Tsallis distribution function and Weibull function are calculated using equations (3,4,5) and (7,8) and fitted to the experimental data. Figure 1 shows the Tsallis fits to the data and figure 2 shows the Weibull distributions fitted to the data in  different rapidity intervals at various energies. Both Tsallis and Weibull functions are also fitted to the data in the full phase space and the results are shown in figure 3. The fit procedure uses ROOT 5.36 from CERN to minimise the $\chi^{2}$ using MINUIT.  

We find that though Weibull gives good fitting in restricted rapidity intervals but fails to reproduce the distributions in the high rapidity intervals and also in full phase space. While Tsallis distribution shows good fits in both full phase space and separately in each rapidity interval. Detailed comparison between the two functions is shown in Table I where $\chi^{2}/ndf$ at all energies for all rapidity intervals are compared. 
It is observed that the  $\chi^{2}/ndf$ values are several orders smaller for the Tsallis fittings as compared to the Weibull fittings. This is true for all rapidity intervals as well as for the full phase space. 

\begin{figure}
\includegraphics[width=3.8 in]{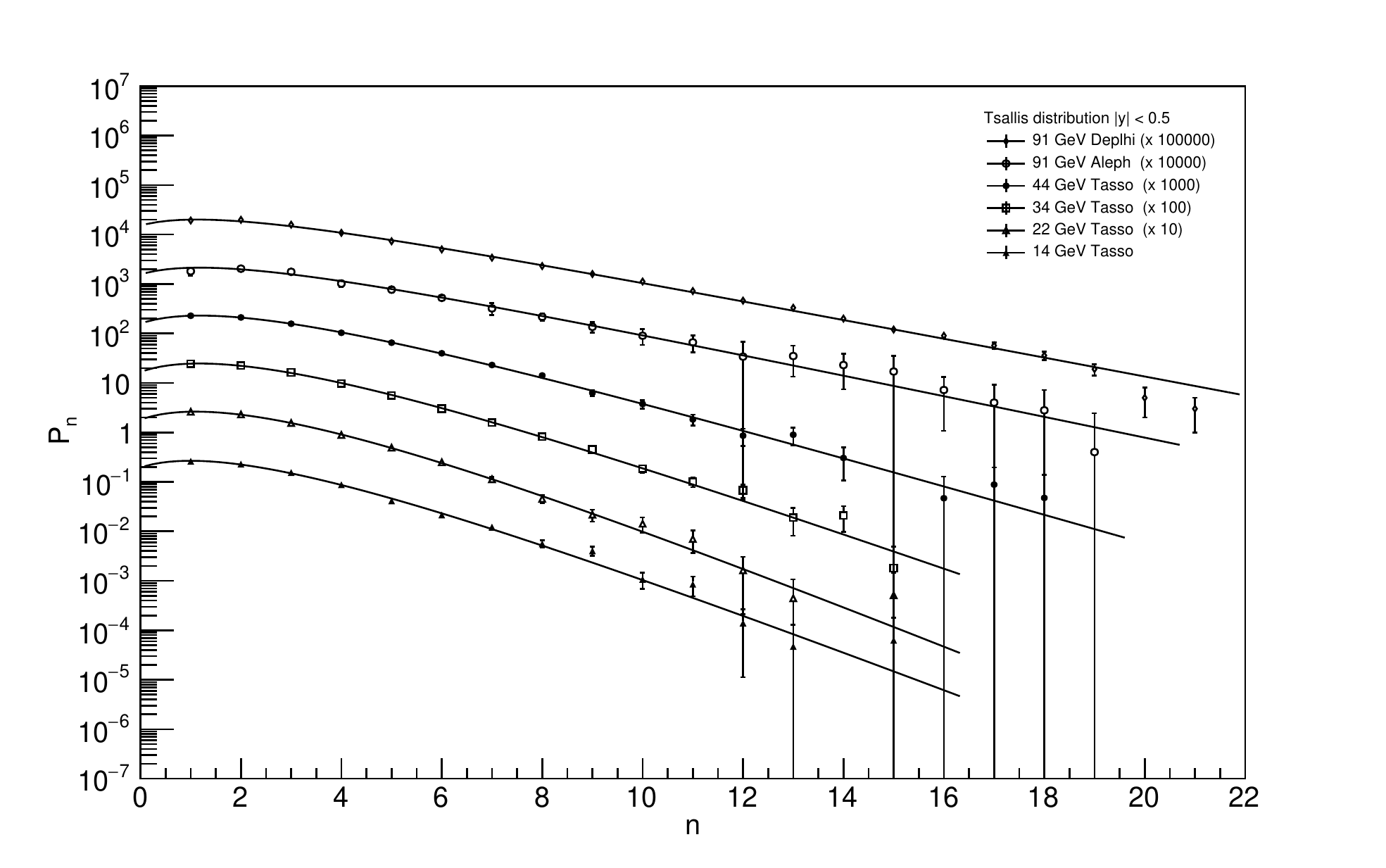}
\includegraphics[width=3.8 in]{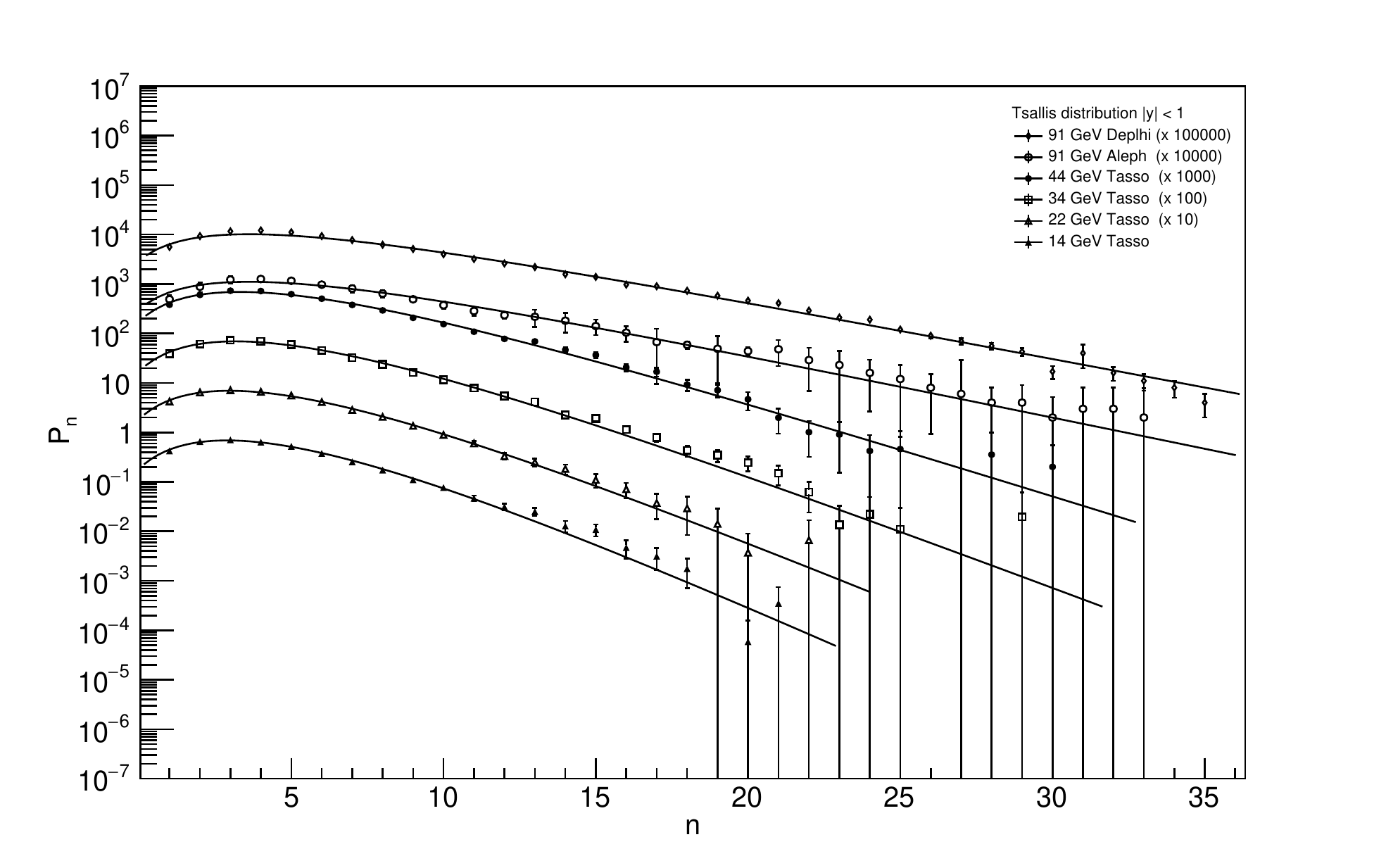}
\includegraphics[width=3.8 in]{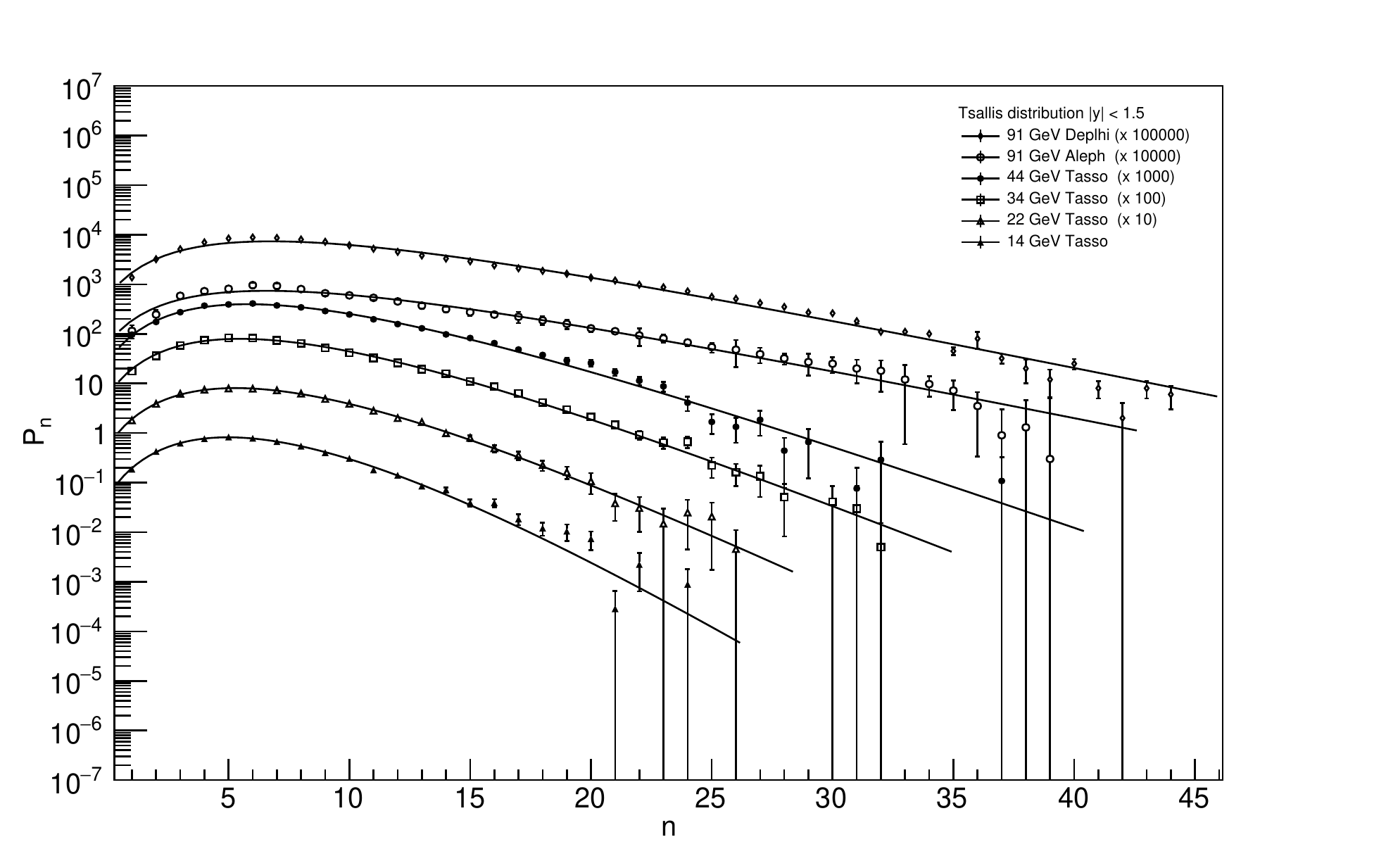}
\includegraphics[width=3.8 in]{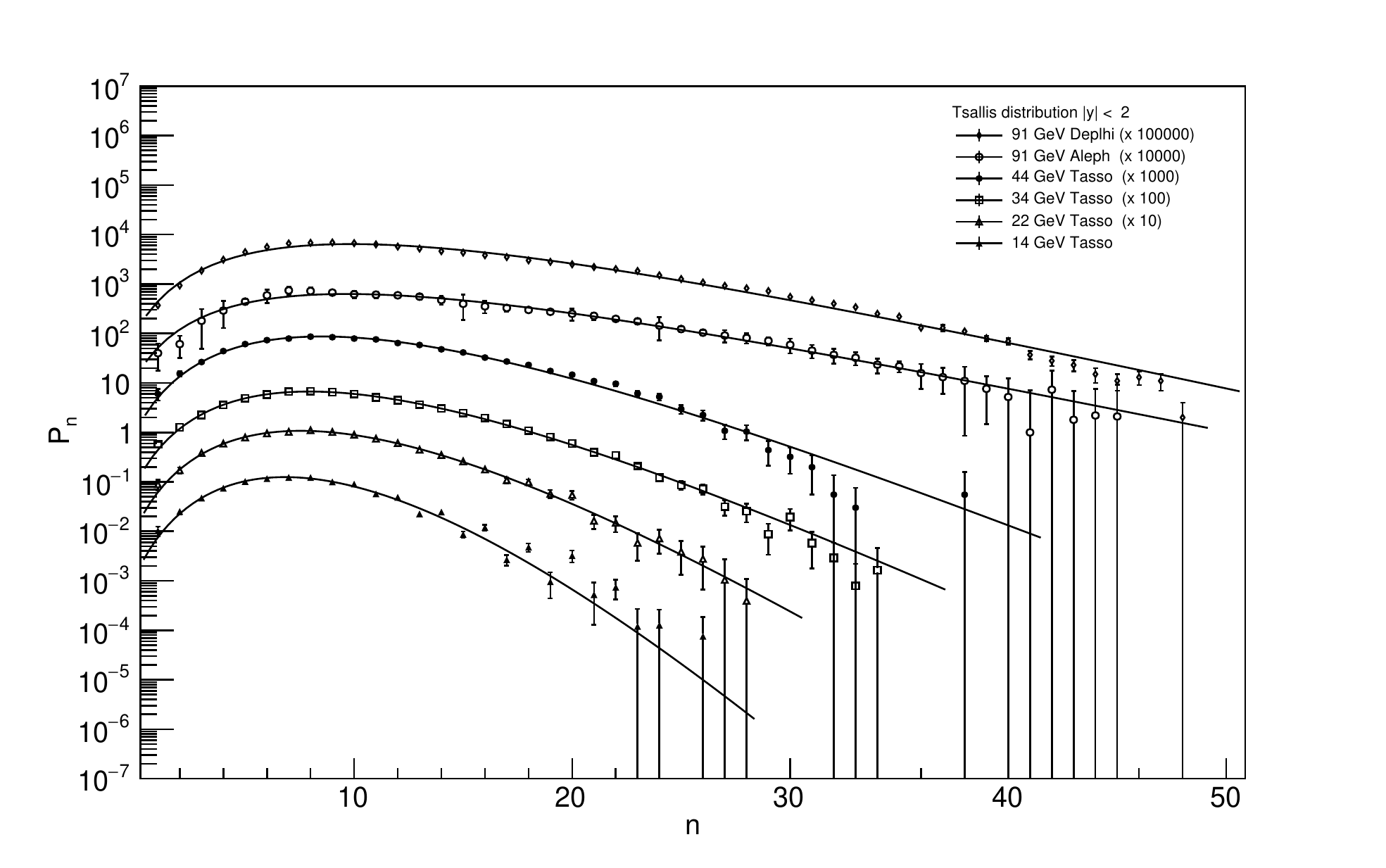}

\caption{Charged multiplicity distribution from top to bottom, $|y|<0.5$, $|y|<1$, $|y|<1.5$ and $|y|<2$ at $\sqrt{s}$=14, 22, 34.8, 44 and 91 GeV. Distributions are multiplied by a factor of 100 for each successive curve.~Solid lines represent the Tsallis distribution and points represent the data}
\end{figure}

\begin{figure}
\includegraphics[width=3.7 in]{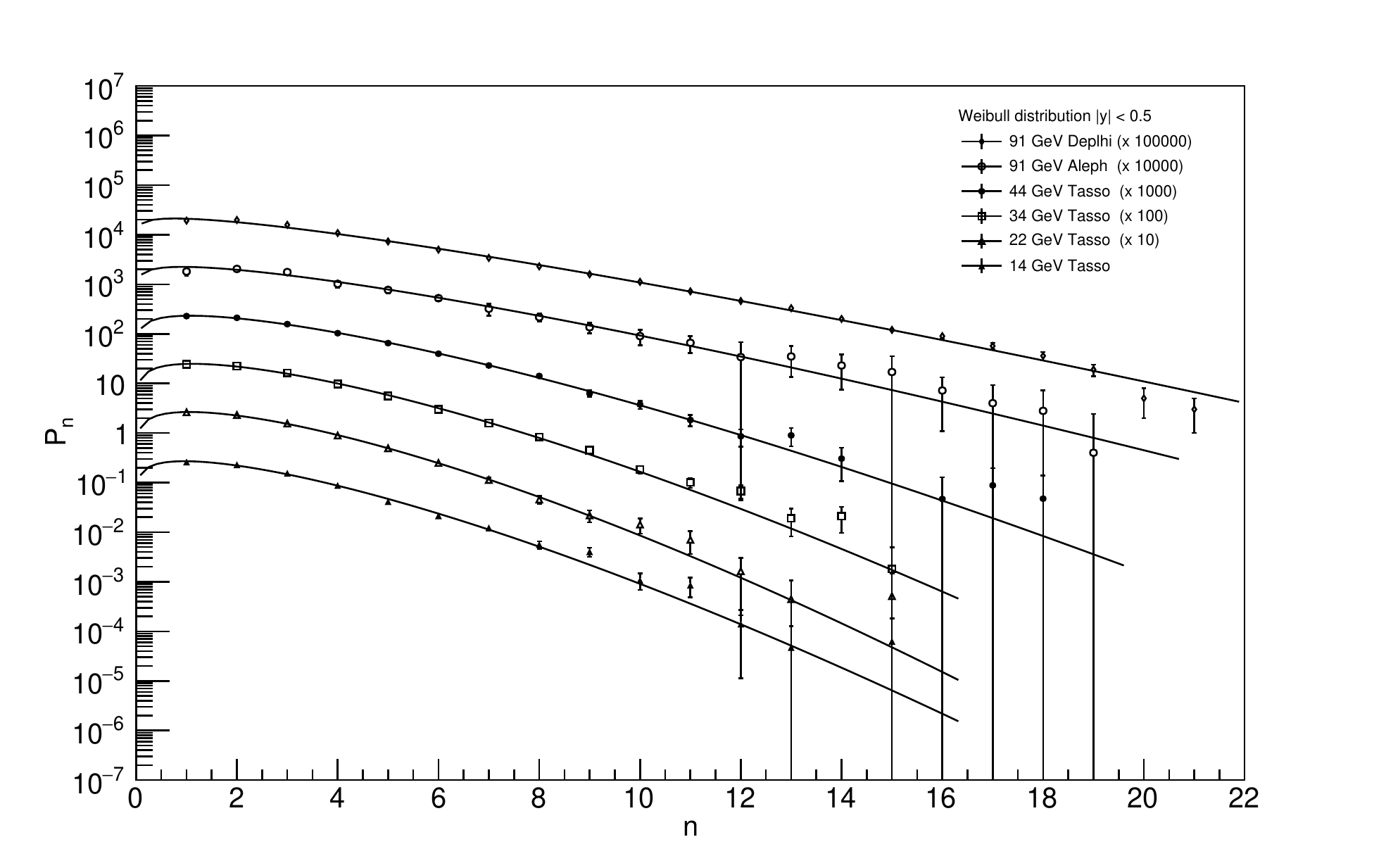}
\includegraphics[width=3.7 in]{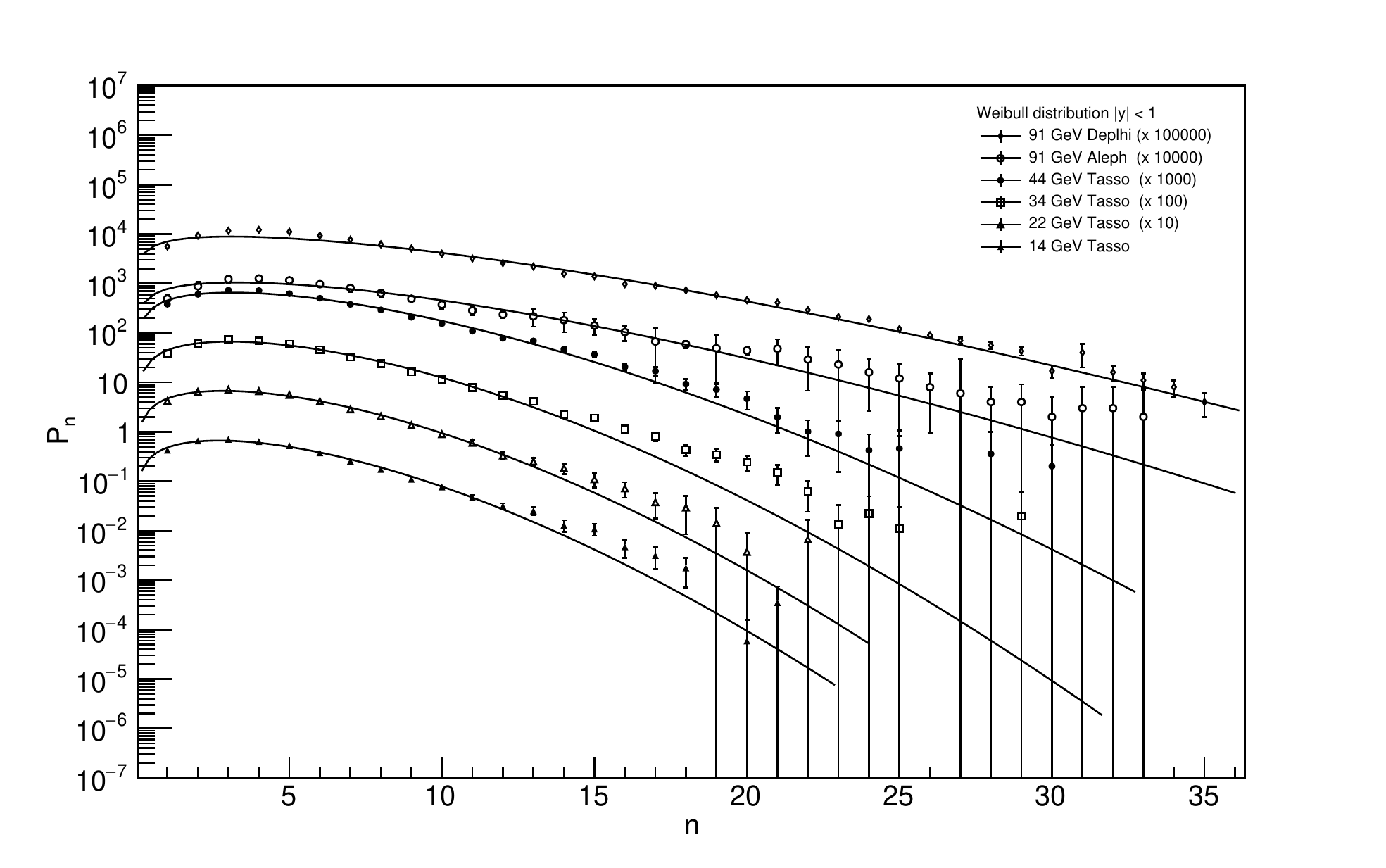}
\includegraphics[width=3.7 in]{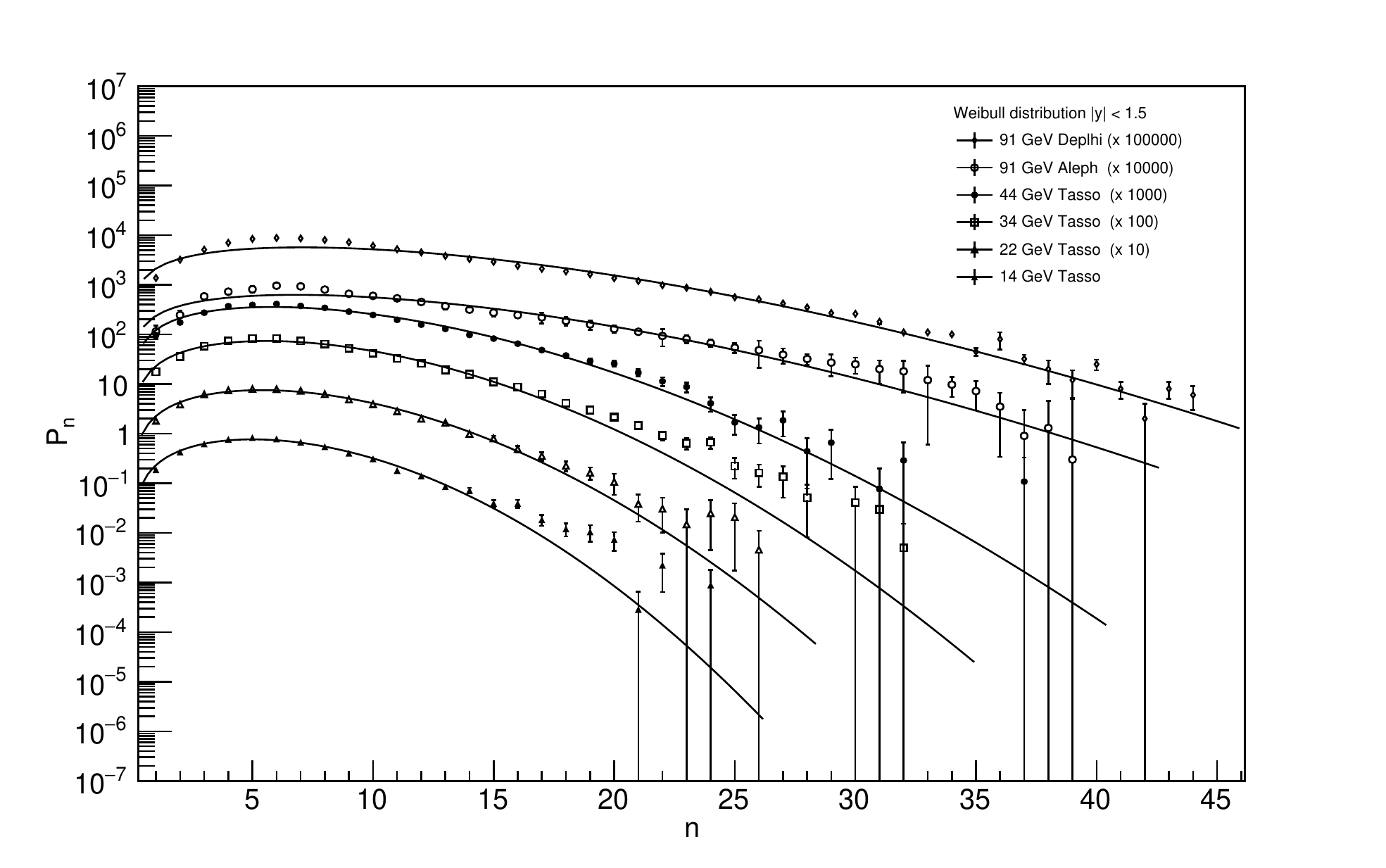}
\includegraphics[width=3.7 in]{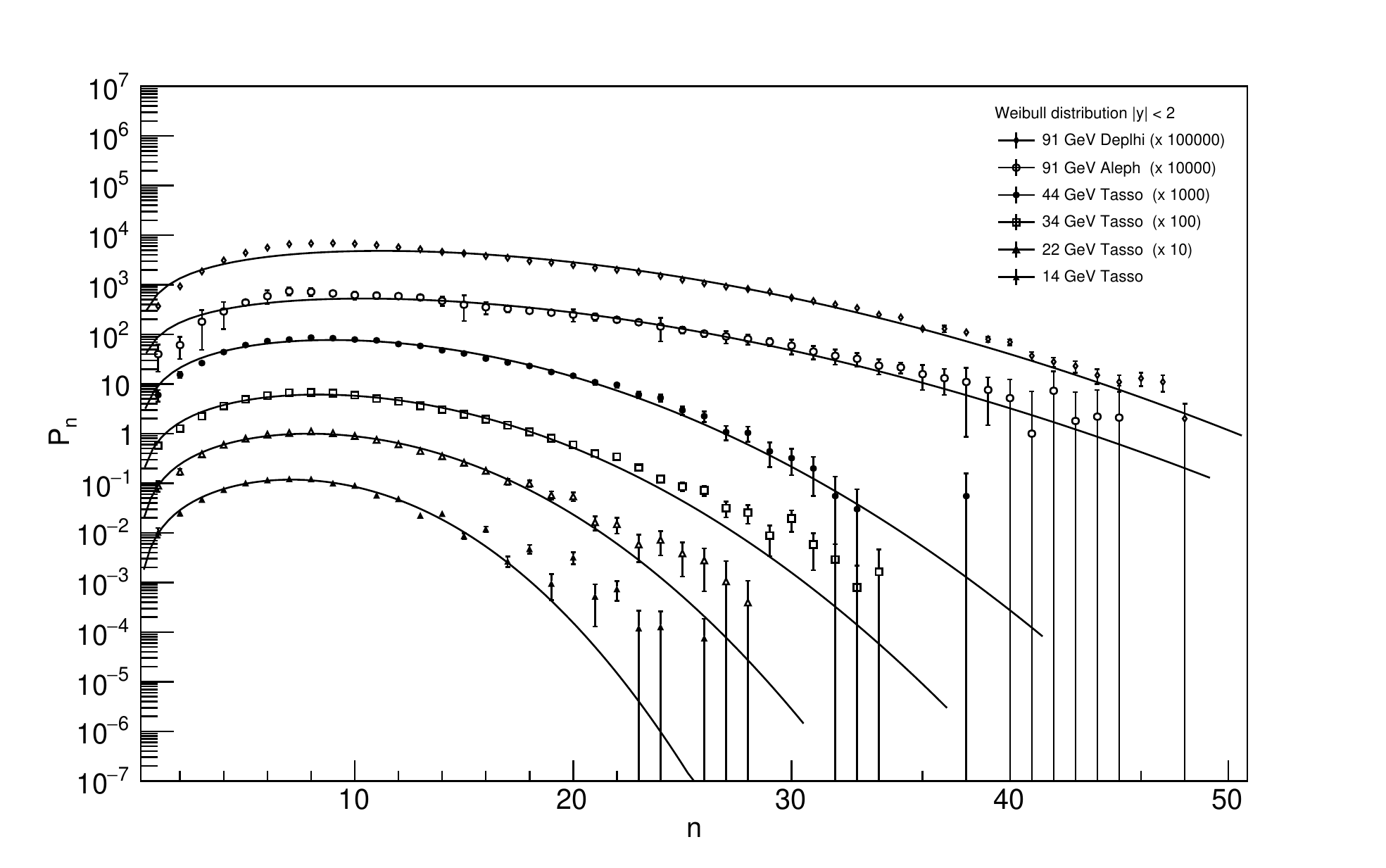}
\caption{Charged multiplicity distributions from top to bottom for $|y|<0.5$, $|y|<1$, $|y|<1.5$ and $|y|<2$ at $\sqrt{s}$=14, 22, 34.8, 44 and 91 GeV. Distributions are multiplied by a factor of 100 for each successive curve. Solid lines represent the Weibull distributions fitted to the data.}
\end{figure}
 
Table~II and Table~III give parameters of both Weibull and Tsallis distributions for extreme rapidity intervals, $|y|< 0.5$ and $|y|< 2$. To avoid too many tables, we are not including the parameter values for other rapidity intervals. We also show in Table IV, the parameter values for full phase space for both functions. 
A comparison of the values in Table I reveals that $\chi^{2}/ndf$ values become worse as we go from lower rapidity to higher rapidity range in both the distributions. However the $\chi^{2}/ndf$ values for the Tsallis distributions are again lower by several orders, confirming that Tsallis distribution fits the data far better than Weibull.

From Tables II, III \& IV, we also observe that for Weibull distribution, as expected, $\lambda$ values increase with energy as well as with rapidity. Similarly for Tsallis distribution, the $q$ value which measures the entropic index of the Tsallis statistics, increases with energy and is more than 1 in every case. This confirms that Tsallis statistics becomes non-extensive.

 \subsection{Approach II} 
It was observed that the multiplicity distributions have a shoulder-like structure at high energies \cite{Gov}. Thus the Tsallis and Weibull distributions both give very high $\chi^{2}/ndf$ values and do not describe the data well at high energy. In our previous publication \cite{SS} we suggested to adopt the Giovannini's \cite{Gov} approach whereby the multiplicity distribution is obtained by using the weighted superposition of two distributions; one accounting for the 2-jet events and another for multi-jet events. For the present work, we use this approach on both Tsallis and Weibull distributions. We call these as modified Tsallis and modified Weibull distributions. The probability functions are  given in equations 6 $\&$ 9. The $\alpha$ in the two equations is the 
 2- jet fraction derived from the DURHAM algorithm, as explained in \cite{DURHAM}, \cite{SD}.\\ 
\begin{figure}[h]
\includegraphics[width=3.5 in]{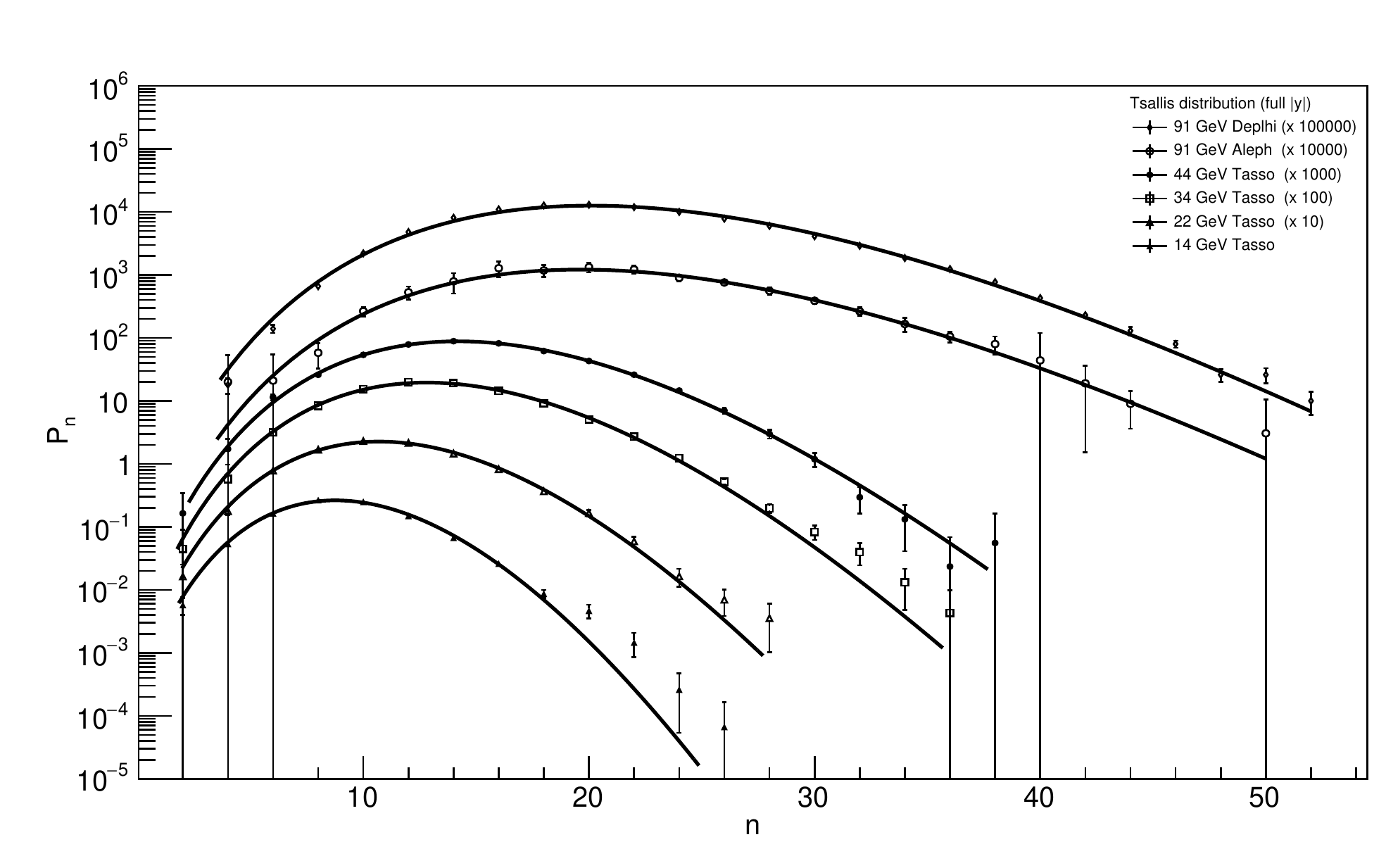}
\includegraphics[width=3.5 in]{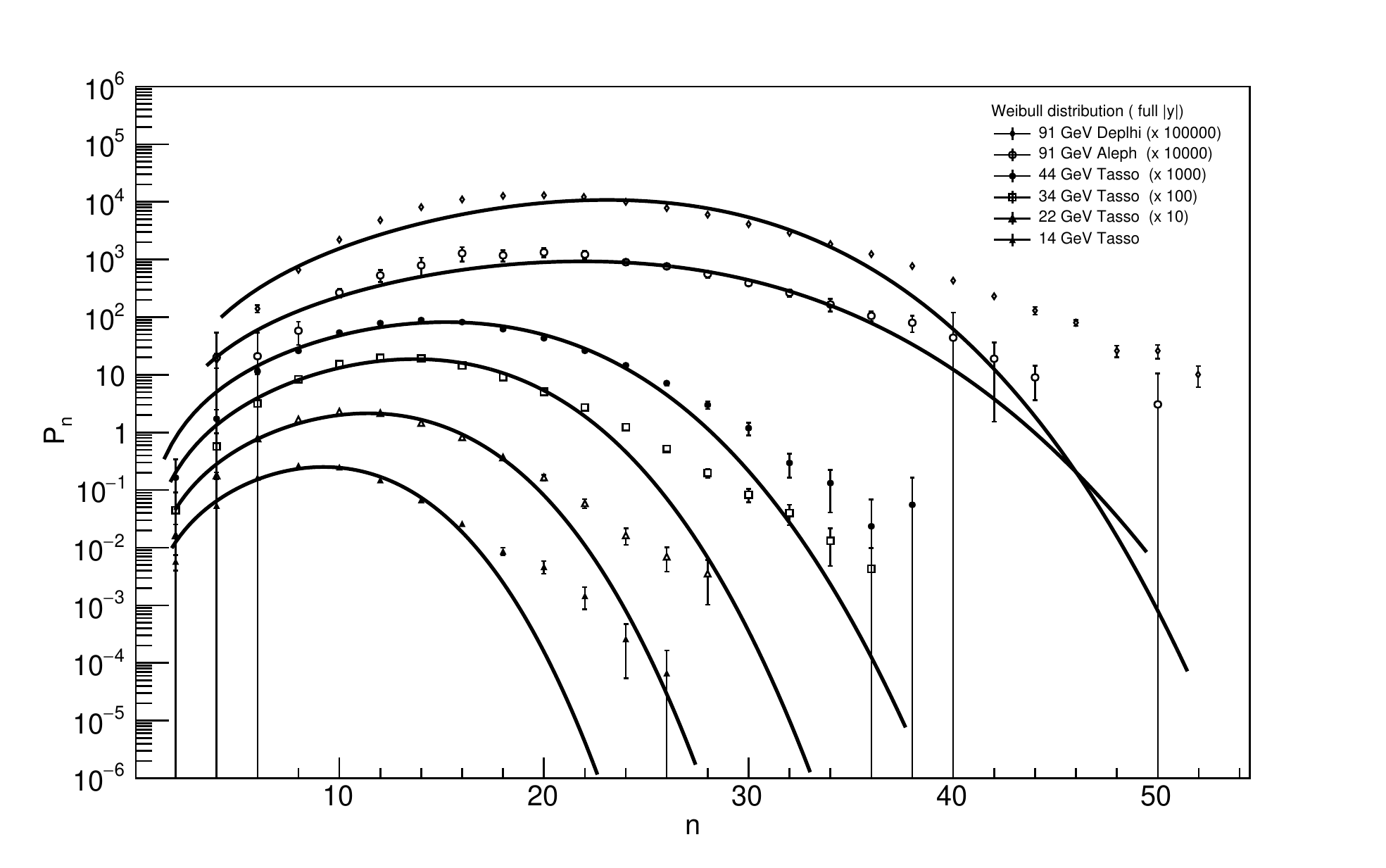}
\caption{Charged multiplicity distribution for full phase space at $\sqrt{s}$=14, 22, 34.8, 44 and 91 GeV. Distributions are multiplied by a factor of 100 for each successive curve. Solid lines represent the Tsallis distribution(up) and Weibull distribution(down) fitted to the data}
\end{figure} 

\begin{figure}[h]
\includegraphics[width=3.2 in]{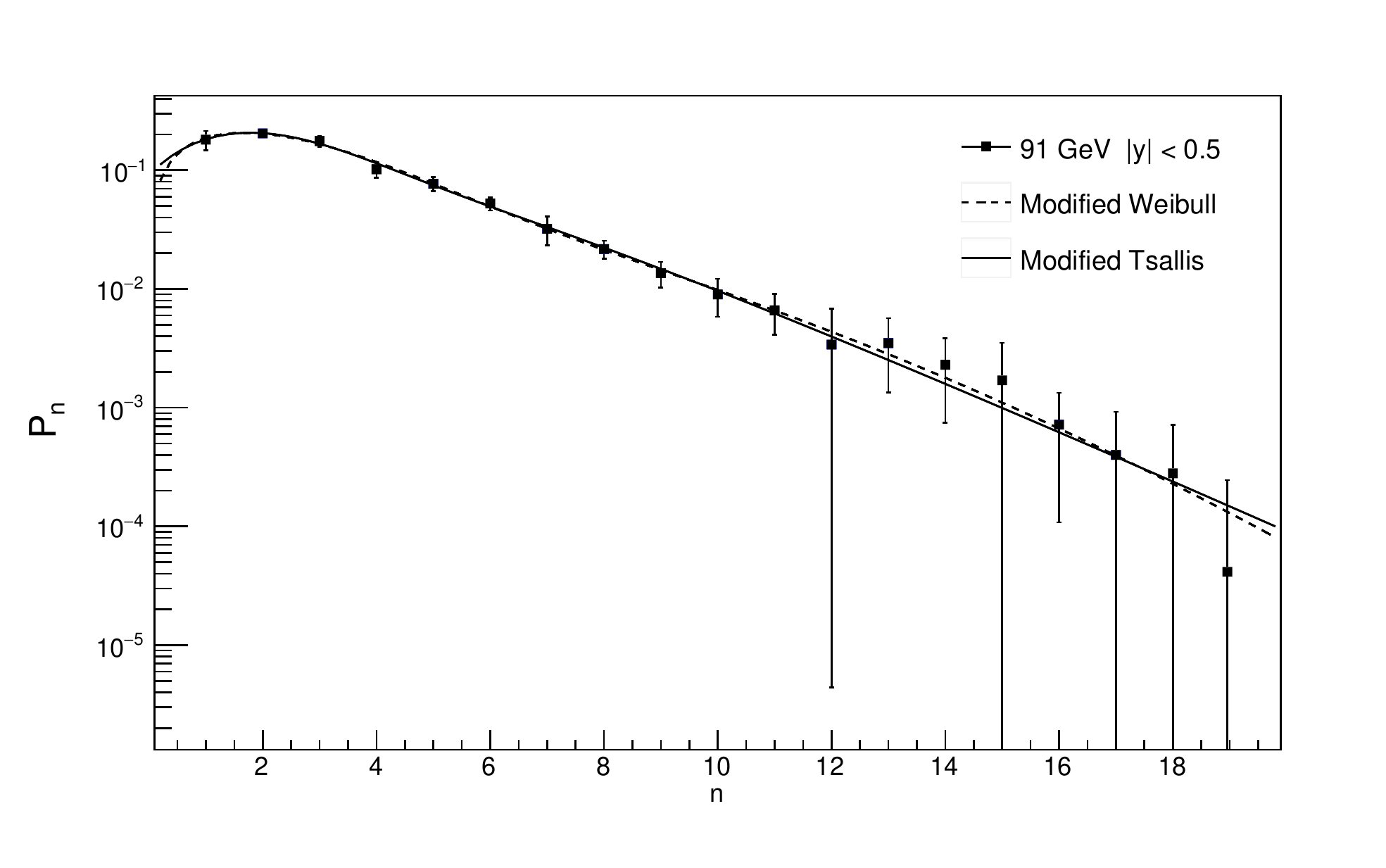}
\includegraphics[width=3.2 in]{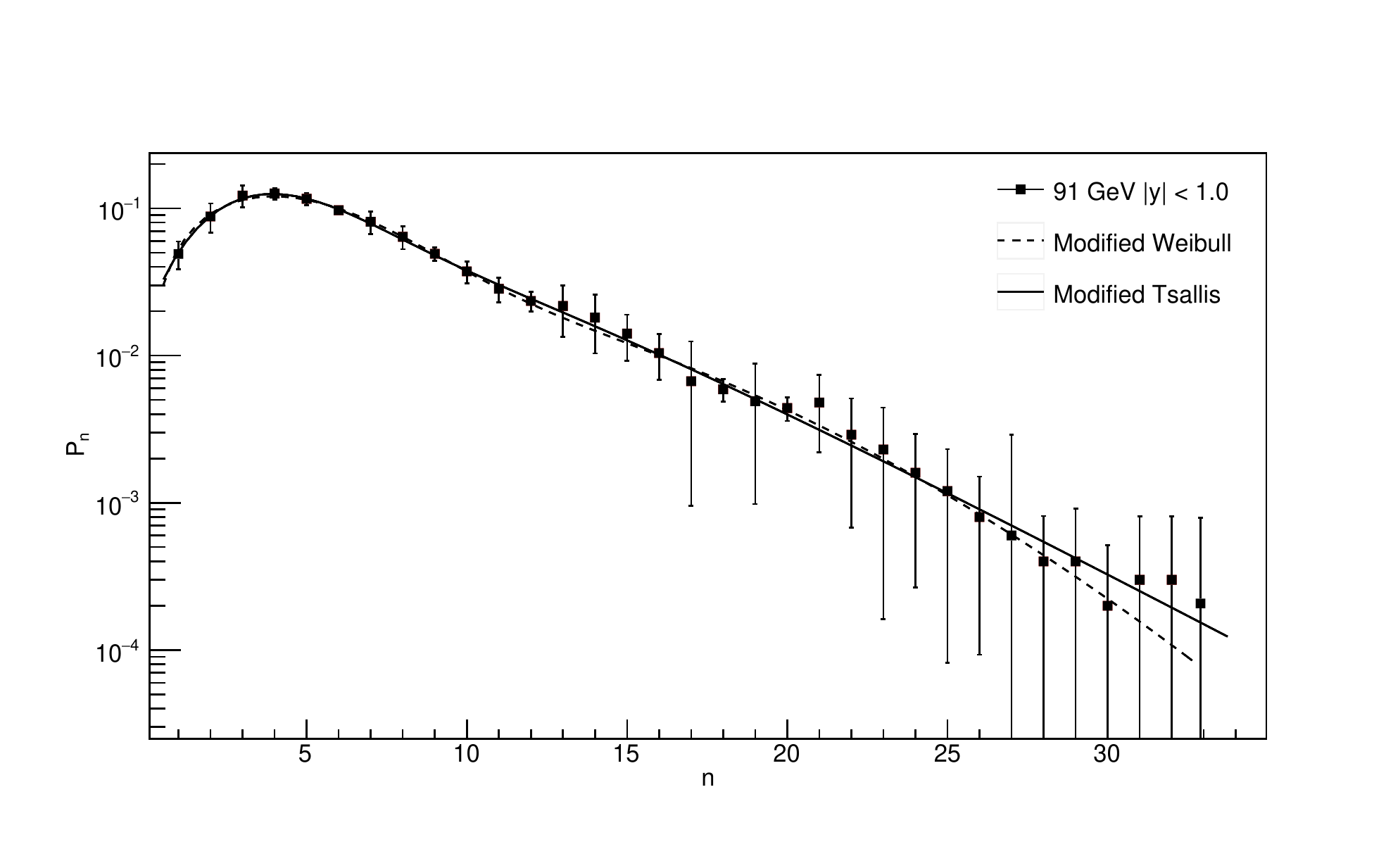}
\includegraphics[width=3.2 in]{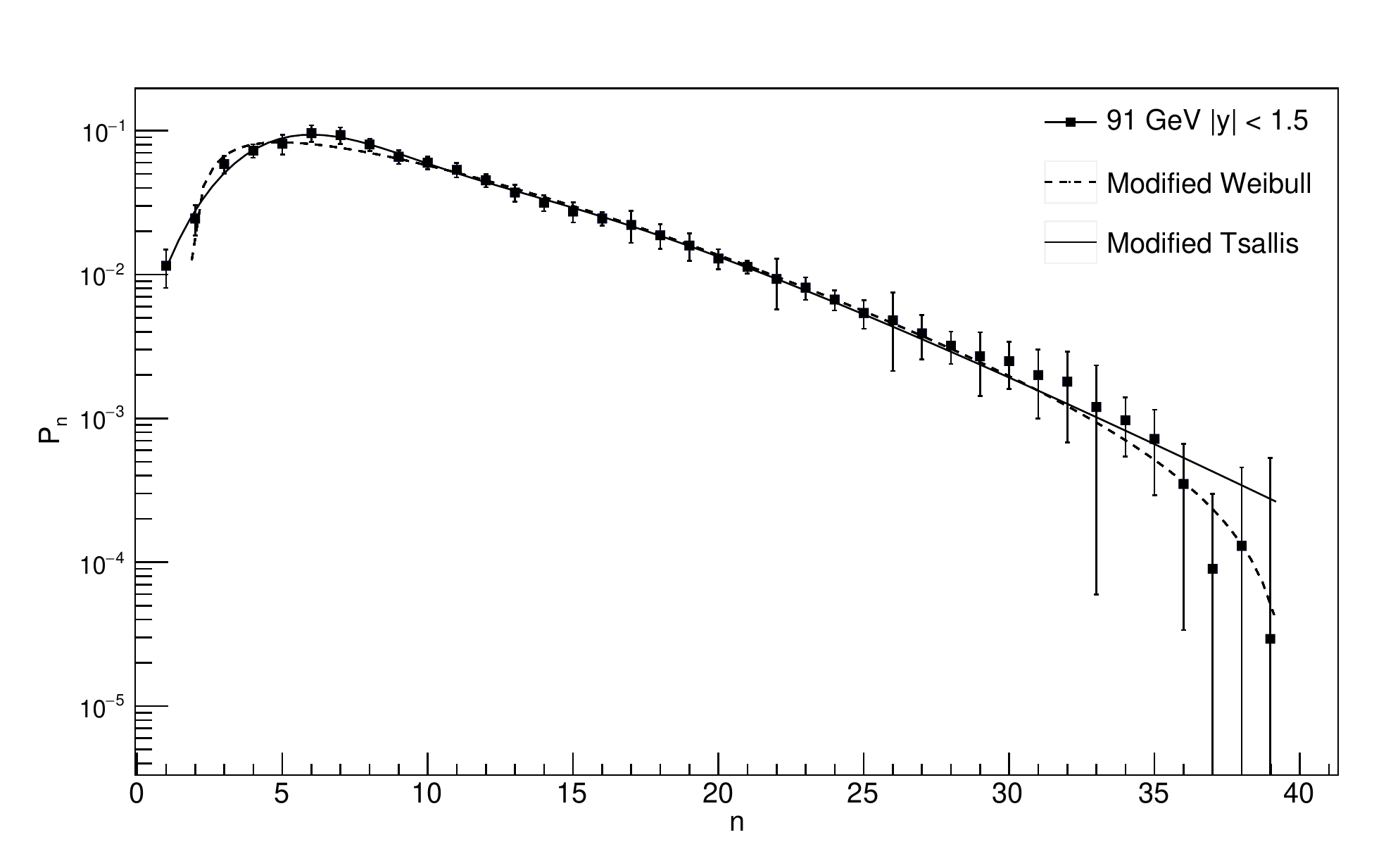}
\includegraphics[width=3.2 in]{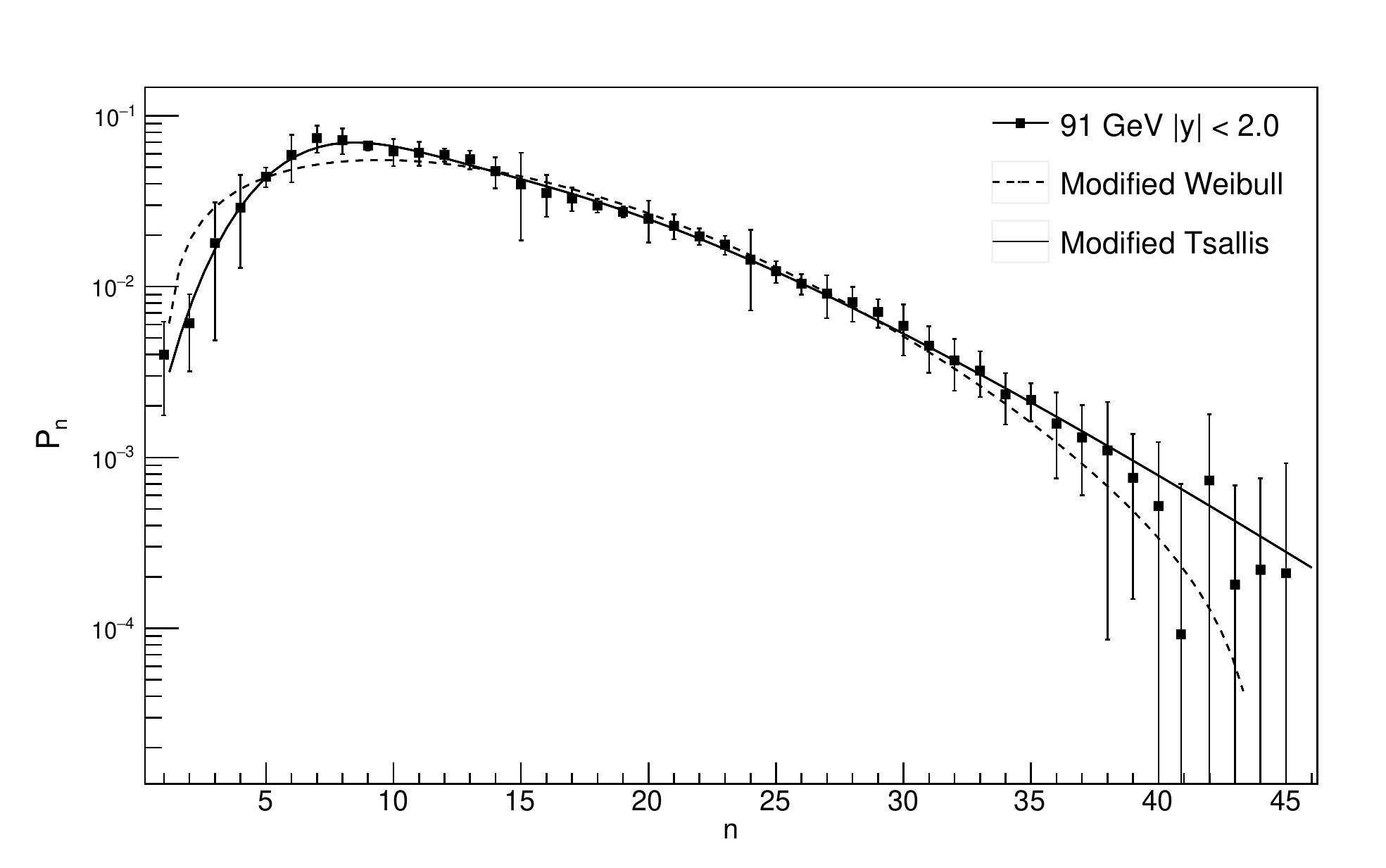}
\includegraphics[width=3.2 in]{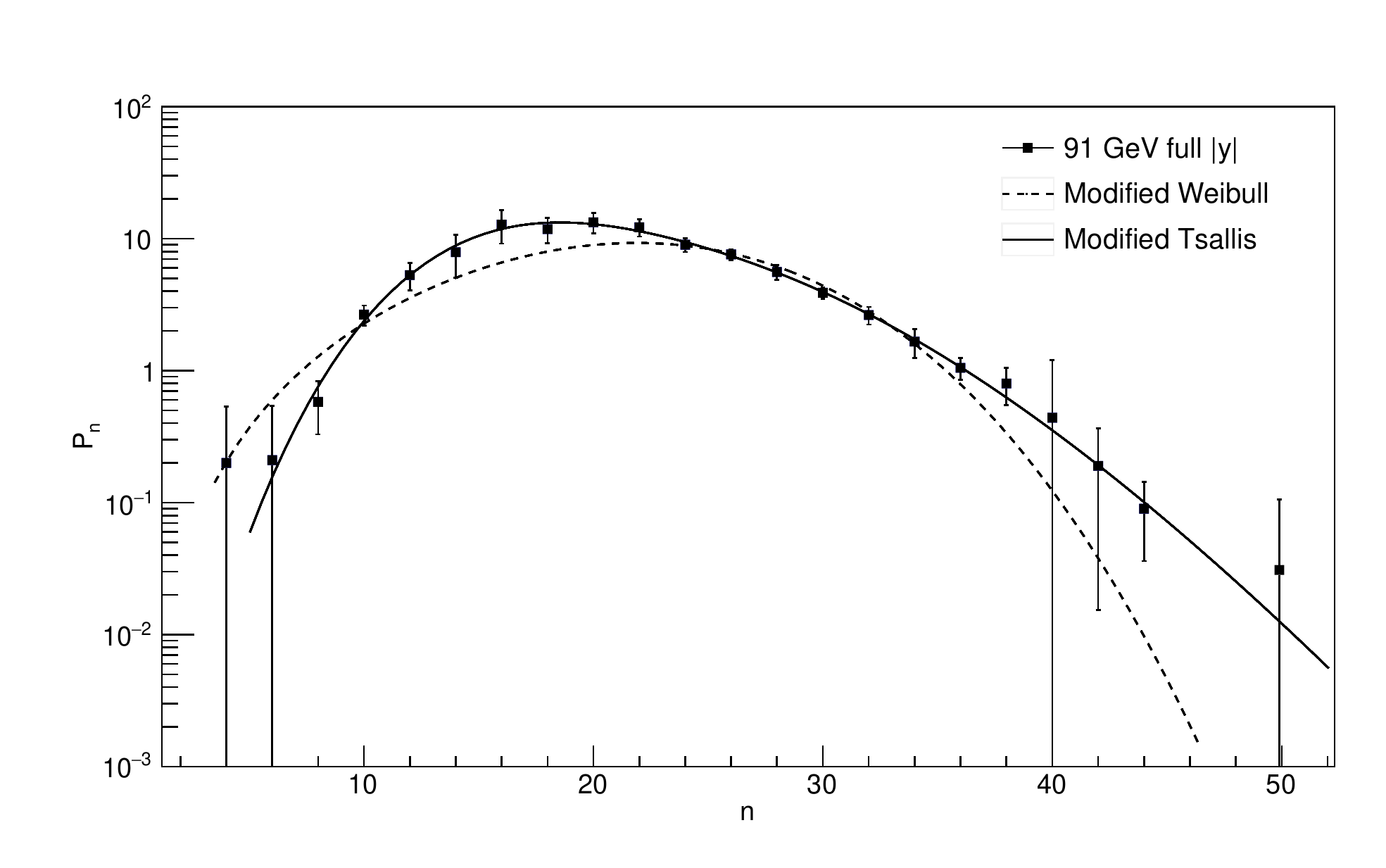}
\caption{Multiplicity distributions at $\sqrt{s}= 91$ GeV for the modified Tsallis distribution and the modified Weibull distribution for $|y|<0.5$, $|y|<1.0$, $|y|<1.5$, $|y|<2.0$ and full phase space from top to bottom}.
\end{figure}

Results using approach II for $\sqrt{s}$= 91 GeV are shown in the figure 4 for various rapidity intervals as well as in the phase space. We use this data as the shoulder structure is prominent at this energy. It can be observed that Tsallis clearly gives an excellent fitting while Weibull fails, particularly at high rapidity range and in full phase space. Comparison of $\chi^{2}/ndf$ values for both Weibull and Tsallis distributions in the four rapidity intervals and full phase space are given in Table V. It can be observed from figure 4 and table V that by using this approach, the fits to the data improve enormously and the $\chi^{2}/ndf$ values decrease substantially. Modified Weibull distribution still remains off data at higher multiplicity values and high rapidity intervals but modified Tsallis distribution fits excellently well. Thus Weibull function has limited applicability, both in its regular form as well as in modified form.
\\
\\
\section{CONCLUSION}
Detailed analysis of the data on $e^{+}e^{-}$ collisions at energies $\sqrt{s}$=14 to 91
GeV has been done by considering the recently proposed Weibull distribution in comparison
to the Tsallis distribution. It is observed that both Weibull and Tsallis distributions successfully explain the multiplicity distributions in restricted rapidity intervals. However the Weibull distribution fails to explain the data in the large rapidity intervals and for full phase space while the Tsallis distribution gives good results.\\ 
It is known that the multiplicity distributions at higher energies show a shoulder structure. In order to improve upon the fits to the data, we propose to build the multiplicity distribution by a convolution of 2-jet component and the multi-jet component by appropriately weighting with the 2-jet fraction obtained from DURHAM algorithm for $\sqrt{s}$ = 91 GeV. By doing so we have shown that the Tsallis distribution explains the data  excellently well, giving the best possible $\chi^{2}/ndf$ values and Weibull distribution though improves the fits by several orders, fails in full phase space. The $q$ value known as entropic index for the Tsallis distribution is greater than 1 in every distribution. This confirms that the Tsallis $q$-statistics has non-extensive behaviour of entropy in both restricted rapidity intervals as well as in full phase space.\\ 
\begin{table*}[t]
\begin{tabular}{|c|ccccc|ccccc|c|}
\hline
 & Weibull& &&& & Tsallis &&&&& \\ 
 & $\chi^{2}/ndf \rightarrow$ & &&&&& &&&& Reference \\\hline 
 Energy & $|y|< 0.5$ & $|y|< 1$ & $|y|< 1.5$ & $|y|< 2$ & full $|y|$ & $|y|< 0.5$ & $|y|< 1$ & $|y|< 1.5$ & $|y|< 2$ & full $|y|$ & \\
(GeV)&       & 		 & 		& 		& 		& 		& 		& 		& 	   &   	& \\\hline  
14   & 1.594 & 2.371 & 3.127 & 4.632 & 10.56 & 1.158 & 1.360 & 1.664 & 3.453 & 2.23 & \cite{TASSO}\\\hline
 22 & 0.391 & 1.481 & 1.562 & 2.042 & 11.15 & 0.273 & 0.466 & 0.288 & 0.505 & 0.904 & " \\\hline
 34.8 & 2.160 & 8.782 & 9.903 & 6.902 & 40.746  & 0.932 & 3.088 & 1.753 & 1.044 & 2.445 & "  \\\hline
 44 & 0.381 & 3.235 & 2.936 & 2.412 & 9.256  & 0.289 & 1.255 & 1.388 & 1.534 & 0.453  & " \\\hline
91 & 0.393 & 1.112 & 1.850 & 1.464 & 1.831 & 0.276 & 0.540 & 0.823 & 0.386 & 0.273 & \cite{ALEPH}\\\hline
91 & 2.107 & 9.197 & 15.395 & 16.813 & 69.273 & 1.990 & 4.034 & 6.515 & 5.520 & 3.795 & \cite{DELPHI}\\\hline
\end{tabular}
\caption{$\chi^{2}/ndf$ comparison for all rapidity intervals and full phase space of Weibull and Tsallis distributions.} 
\end{table*}
\begin{table*}[t]
\begin{tabular}{|c|c|c|c|c|c|c|c|c|}
\hline
  &         &  			  & & &          &                &     &\\
  & Weibull &$\rightarrow$ & & & Tsallis  &$\rightarrow$   &     & \\\hline       
  Energy & k  & $\lambda$ & $\chi^{2}/ndf$ & $nV$ & $nv_{0}$ & q & $\chi^{2}/ndf$ & Reference\\
  (GeV)	 &	  &			 &				&	 &	  &   &    &   \\\hline           
 14 & 1.37 $\pm$ 0.030 & 2.46 $\pm$ 0.043 & 1.594 & 1.22 $\pm$ 0.067 & -0.146 $\pm$ 0.008 & 1.262 $\pm$ 0.087 & 1.158 & \cite{TASSO}\\\hline
 22 &  1.41 $\pm$ 0.031 & 2.53 $\pm$ 0.046 & 0.391 & 1.31 $\pm$ 0.071 & -0.138 $\pm$ 0.008 & 1.178 $\pm$ 0.062 & 0.273 & " \\\hline
34.8 &  1.40 $\pm$ 0.014 & 2.72 $\pm$ 0.02 & 2.160 & 1.31 $\pm$ 0.033 & -0.158 $\pm$ 0.004 & 1.282 $\pm$ 0.042 & 0.932 & " \\\hline
44 &  1.35 $\pm$ 0.023 & 2.95 $\pm$ 0.043 & 0.381 & 1.25 $\pm$ 0.054 & -0.186$\pm$ 0.005 & 1.212 $\pm$ 0.082 & 0.289 & " \\\hline
91 &  1.25 $\pm$ 0.074 & 3.31 $\pm$ 0.166 & 0.393 & 1.14 $\pm$ 0.167 & -0.227 $\pm$ 0.013 & 2.845 $\pm$ 1.373 & 0.276 & \cite{ALEPH}\\\hline
 91 &  1.20 $\pm$ 0.013 & 3.39 $\pm$ 0.05 & 2.107 & 1.08 $\pm$ 0.036 & -0.238 $\pm$ 0.002 & 3.696 $\pm$ 0.390 & 1.990 & \cite{DELPHI} \\\hline

\end{tabular}
\caption{Parameters of Weibull and Tsallis functions for $|y|<0.5$ }  
\end{table*}
\begin{table*}[t]
\small
\begin{tabular}{|c|c|c|c|c|c|c|c|c|}
\hline
  &         &  			  & & &          &                &     &\\
  & Weibull &$\rightarrow$ & & & Tsallis  &$\rightarrow$   &     & \\\hline       
  Energy & k  & $\lambda$ & $\chi^{2}/ndf$ & $nV$ & $nv_{0}$ & q & $\chi^{2}/ndf$ & Reference\\
  (GeV)	 &	  &			 &				&	 &	  &   &    &   \\\hline    
14 & 2.58 $\pm$ 0.035 & 8.58 $\pm$ 0.052 & 4.632 & 5.64 $\pm$ 0.145 & -0.119 $\pm$ 0.008 & 1.0012 $\pm$ 0.0009 & 3.453& \cite{TASSO}\\\hline
 22 &  2.44 $\pm$ 0.033 & 9.82 $\pm$ 0.066 & 2.042 & 5.19 $\pm$ 0.124 & -0.180 $\pm$ 0.005 & 1.0127 $\pm$ 0.0046 & 0.505& " \\\hline
34.8 &  2.28 $\pm$ 0.015 & 10.72 $\pm$ 0.037 & 6.902 & 4.68 $\pm$ 0.050 & -0.221 $\pm$ 0.002 & 1.0556 $\pm$ 0.0046 & 1.044& "\\\hline
44 &  2.21 $\pm$ 0.023 & 11.77 $\pm$ 0.073 & 2.412 & 4.64 $\pm$ 0.082 & -0.237$\pm$ 0.002 & 1.0994 $\pm$ 0.0157 & 1.534& "\\\hline
91 &  1.96 $\pm$ 0.039 &  14.96 $\pm$ 0.220 & 1.464 & 3.71 $\pm$ 0.146 & -0.296 $\pm$ 0.003 & 2.1550 $\pm$ 0.2500 & 0.386& \cite{ALEPH}\\\hline
91 &  2.02 $\pm$ 0.009 & 15.73 $\pm$ 0.053 & 16.813 & 3.91 $\pm$ 0.029 & -0.292 $\pm$ 0.001 & 1.892 $\pm$ 0.0375 & 5.520& \cite{DELPHI}\\\hline
\end{tabular}
\caption{Parameters of Weibull and Tsallis functions for $|y|<2.0$ }  
\end{table*}
\begin{table*}[t]
\small
\begin{tabular}{|c|c|c|c|c|c|c|c|c|}
\hline
  &         &  			  & & &          &                &     &\\
  & Weibull &$\rightarrow$ & & & Tsallis  &$\rightarrow$   &     & \\\hline       
  Energy & k  & $\lambda$ & $\chi^{2}/ndf$ & $nV$ & $nv_{0}$ & q & $\chi^{2}/ndf$ & Reference\\
  (GeV)	 &	  &			 &				&	 &	  &   &    &   \\\hline    
14 & 3.42 $\pm$ 0.049 & 10.22 $\pm$ 0.055 & 10.56 & 9.40 $\pm$ 0.280 & -0.008 $\pm$ 0.0014 & 1.00001 $\pm$ 0.00002 & 2.23&\cite{TASSO}\\\hline
 22 &  3.61 $\pm$ 0.050 & 12.49 $\pm$ 0.069 & 11.15 & 10.12 $\pm$ 0.260 & -0.049 $\pm$ 0.011 & 1.00001 $\pm$ 0.00002 & 0.904& " \\\hline
 34.8 &  3.73 $\pm$ 0.024 & 14.87 $\pm$ 0.036 & 40.746 & 10.81 $\pm$ 0.158 & -0.095 $\pm$ 0.004 & 1.00011 $\pm$ 0.00007 & 2.445& " \\\hline

 44 &  3.62 $\pm$ 0.040 & 16.64 $\pm$ 0.072 & 9.256 & 10.91 $\pm$ 0.219 & -0.128$\pm$ 0.006 & 1.00055 $\pm$ 0.00036 & 0.453 & " \\\hline
91 &  3.67 $\pm$ 0.114 & 23.87 $\pm$ 0.284 & 1.831 & 11.30 $\pm$ 0.530 & -0.209 $\pm$ 0.010 & 1.01136 $\pm$ 0.00730 & 0.273 & \cite{ALEPH}\\\hline
91 &  4.21 $\pm$ 0.025 & 24.63 $\pm$ 0.076 & 69.273 & 11.72 $\pm$ 0.010 & -0.206 $\pm$ 0.002 & 1.00960 $\pm$ 0.00120 & 3.795& \cite{DELPHI}\\\hline
\end{tabular}
\caption{Parameters of Weibull and Tsallis function for full rapidity window.}
\end{table*}
\begin{table*}[b]
\begin{tabular}{|c|c|c|c|c|}
\hline
	   &                      &                     &                   &                  \\
$|y|$  & Weibull Distribution & Tsallis Distribution & Modified Weibull  & Modified  Tsallis\\
        &$\chi^{2}/ndf \rightarrow $ &&&               \\\hline
0.5        & 0.395 & 0.276 & 0.162 & 0.164 \\\hline
1          & 1.112 & 0.540 & 0.099 & 0.075 \\\hline
1.5        & 1.843 & 0.823 & 1.101 & 0.209 \\\hline
2          & 1.463 & 0.386 & 1.115 & 0.122 \\\hline
full $|y|$ & 1.831 & 0.272 & 2.041 & 0.165 \\\hline
\end{tabular}
\caption{$\chi^{2}/ndf$ comparison for Weibull, Tsallis, Modified Weibull and Modified Tsallis distributions at $\sqrt{s}$= 91 GeV (ALEPH)for all rapidity intervals} 
\end{table*}

\end{document}